\documentclass[letterpaper,twocolumn,10pt]{article}

\usepackage{newtxtext}
\usepackage{amssymb}
\usepackage{newtxmath}
\usepackage{helvet}
\usepackage{courier}

\usepackage[margin=0.75in,top=1.25in,bottom=1.25in,nohead]{geometry}

\usepackage[hyphens]{url}
\usepackage{graphicx}
\usepackage{natbib}
\usepackage{caption}
\usepackage{amsmath}
\usepackage{booktabs}
\usepackage{multirow}

\title{\bfseries Toward Annotation-Efficient Continuous Emotion Arousal Quantification via Group-Level EEG Dynamic Neural Synchrony}

\author{
    Guandong Pan\textsuperscript{1},
    Yaqian Yang\textsuperscript{2},
    Shi Chen\textsuperscript{2},
    Yi Zheng\textsuperscript{1},
    Yi Zhen\textsuperscript{2},\\[2pt]
    Hongwei Zheng\textsuperscript{3,$\ast$},
    Shaoting Tang\textsuperscript{2,$\ast$}\\[6pt]
    \normalsize
     \textsuperscript{1}Beijing Advanced Innovation Center for Future Blockchain and Privacy Computing,\\
    \qquad Beihang University, Beijing 100191, China\\
    \textsuperscript{2}School of Artificial Intelligence, Beihang University, Beijing 100191, China\\
    \textsuperscript{3}Beijing Academy of Blockchain and Edge Computing, Beijing 100085, China\\[4pt]
    tangshaoting@buaa.edu.cn,~hwzheng@pku.edu.cn
}
\date{}

\begin{document}
\maketitle

\begin{abstract}
Continuous emotional arousal quantification remains bottlenecked by time-consuming and labor-intensive manual annotation. This work investigates group-level EEG dynamic neural synchrony (DNS) as a principled signal for continuous arousal quantification that bypasses per-subject manual labeling. Using Correlated Component Analysis (CorrCA) with sliding-window computation across four EEG datasets spanning 142 subjects and over 207 hours, we systematically evaluate DNS as a group-level marker for emotional arousal dynamics. Three key findings emerge. First, DNS exhibits significant emotion information from valence-dependent differences (all $p{<}0.003$), with positive emotions eliciting higher synchrony. Second, DNS correlates more strongly with the first-order derivative of arousal than with raw arousal values, revealing that neural synchrony captures the rate of emotional change rather than static intensity. Third, we provide the first systematic characterization of how DNS--arousal coupling depends on key methodological choices, finding that moderate windows (10--30\,s), positive lags (0--10 steps), and First-order Difference feature of EEG from the dominant CorrCA component yield consistently strong coupling.  Subject-split replication and block permutation tests confirm these associations are not statistical artifacts. Our findings establish DNS as an empirically validated group-level marker toward annotation-efficient continuous emotional arousal quantification.
\end{abstract}

\section{Introduction}

Continuous emotional arousal quantification is bottlenecked by the time-consuming and labor-intensive  manual annotation. Real-time online annotation introduces dual-task interference and attentional decay~\cite{sharma_dataset_2019,xue_rcea360vr_2021}; post-hoc protocols suffer from memory decay and secondary emotional responses to stimulus re-exposure~\cite{ding_interbrain_2021}. Both derive labels from individual subjective reports, which are noisy, costly to scale, and constrain generalizable supervised models~\cite{du_efficient_2022,baveye_lirisaccede_2015}.

An alternative is to mine group-shared neural signatures that bypass per-subject labeling. Inter-subject correlation (ISC) captures stimulus-driven shared affective processing by quantifying temporal coupling of neural activity across individuals under naturalistic stimulation~\cite{hasson_intersubject_2004,nastase_measuring_2019}. Emotion processing exhibits cross-individual neural consistency~\cite{barrett_theory_2017}, and fMRI studies validate the link between ISC and subjective emotional experience~\cite{nummenmaa_emotions_2012,simony_dynamic_2016}.

Electroencephalogram (EEG) offers millisecond resolution uniquely suited for dynamic emotion analysis~\cite{edelman_noninvasive_2019}, yet EEG-ISC in affective computing remains sparse, focusing on audience preference~\cite{dmochowski_audience_2014} or engagement~\cite{ki_attention_2016,rosenkranz_eegbased_2021}. Existing EEG-ISC work treats synchrony as classifier input rather than investigating its inherent coupling with emotional dynamics~\cite{shen_contrastive_2022,shen_contrastive_2024,pan_realistic_2026}, and relies on fixed short windows (1--5\,s) without empirical optimization~\cite{dmochowski_correlated_2012,ding_interbrain_2021}. Two open questions motivate our work: (1) Does group-level EEG synchrony carry emotion information, and if so, which aspect of emotional dynamics does it reflect? (2) How can we establish principled computational and parametric guidelines for EEG-based group synchrony in emotion research?

We present a systematic empirical investigation. We compute multi-channel dynamic neural synchrony (DNS) via Correlated Component Analysis (CorrCA)~\cite{dmochowski_correlated_2012} across four EEG datasets (SEED, SEED-IV, SEED-VII, and our self-collected BAVE), spanning 142 subjects and over 207 hours. We sweep window size (2--40\,s), time lag ($-$20 to $+$20 steps), feature type (FD, DE across five bands, raw), and CorrCA component (0--4), providing the first comprehensive parameter-level characterization of EEG-ISC for continuous emotion.

Three findings emerge. First, DNS captures emotion-specific information: significant valence-dependent differences appear across datasets ($F{=}10.46$/$F{=}8.19$/$F{=}6.25$, all $p{<}0.003$). Second, DNS correlates more strongly with arousal derivatives than raw values, revealing that neural synchrony captures the rate of emotional change. Third, the DNS-arousal coupling is systematically modulated by parameters: moderate windows (10-30\,s) with positive lags (0-10 steps) and the First-order Difference feature (FD) with Component~0 provide a consistently effective configuration, with subject-split replication and block permutation tests confirming robustness.

These findings establish DNS as a validated population-level proxy for continuous arousal dynamics, providing an annotation-efficient framework that is ready for deployment under naturalistic stimulation paradigms without per-study parameter re-tuning.

\section{Related Work}

\subsection{Inter-Subject Neural Synchrony}

ISC quantifies temporal coupling of neural activity across individuals, capturing group-level shared processing~\cite{hasson_intersubject_2004,hasson_enhanced_2008,nastase_measuring_2019}. Hasson et al.~\cite{hasson_intersubject_2004} demonstrated synchronized brain activity during film viewing; Nummenmaa et al.~\cite{nummenmaa_emotions_2012} showed high-arousal states enhance synchrony; Simony et al.~\cite{simony_dynamic_2016} linked ISC to narrative engagement. Dmochowski et al.~\cite{dmochowski_correlated_2012,dmochowski_audience_2014} pioneered CorrCA for EEG, extending to attention~\cite{ki_attention_2016,rosenkranz_eegbased_2021}. However, these studies target cognitive engagement rather than emotion, and rely on fixed short windows without parameter characterization.

\subsection{Neural Synchrony and Emotion}

Ding et al.~\cite{ding_interbrain_2021} used inter-brain features to predict arousal; Shen et al.~\cite{shen_contrastive_2022,shen_contrastive_2024} proposed CL-SSTER for shared representations. These approaches treat ISC as classifier input rather than investigating its inherent coupling with emotional dynamics. Our work systematically quantifies DNS-arousal coupling across parameter configurations, including window size, temporal alignment, feature type and component order.

\subsection{Annotation Efficiency}

The annotation bottleneck is well-documented~\cite{metallinou_annotation_2013,sharma_continuous_2020}. Existing strategies such as transfer learning~\cite{du_efficient_2022} and domain adaptation~\cite{li_multidomain_2021} reduce labeling requirements on target data, but still presuppose large annotated source datasets. Our work explores a fundamentally complementary direction: DNS is computed purely from neural data alone, independent of manual ratings. The DNS-arousal coupling requires a one-time validation on a calibrated subset; once established, DNS can serve as a surrogate signal for continuous arousal dynamics without per-study labeling.

\section{Methods}

\subsection{Datasets and Preprocessing}

We use four EEG datasets (Table~\ref{tab:dataset_info}). All share 64-channel acquisition (62 effective, 200\,Hz). Continuous arousal: SEED (self-collected, 40 raters/clip), SEED-VII (built-in, 0.25\,Hz), BAVE (self-collected, 20 raters/clip), all resampled to 1\,Hz and averaged. EEG Preprocessing: 200\,Hz downsampling, 0--75\,Hz bandpass, 48--52\,Hz notch, bad channel interpolation, ICA-based artifact removal~\cite{delorme_eeglab_2004}.

\begin{table}[tb]
  \centering
  \caption{Datasets. CA: continuous arousal.}
  \label{tab:dataset_info}
  \resizebox{\columnwidth}{!}{%
  \begin{tabular}{lcccl}
    \toprule
    \textbf{Dataset} & \textbf{Subj.} & \textbf{Ch.} & \textbf{Stimuli} & \textbf{Annotation} \\
    \midrule
    SEED     & 45 (15$\times$3) & 62 & 15 clips, $\sim$4\,min & Discrete (3 cat.) + CA (ours)\\
    SEED-IV  & 15 & 62 & 72 clips, $\sim$2\,min & Discrete (4 cat.) \\
    SEED-VII & 20 & 62 & 80 clips, 2--5\,min & Discrete (7 cat.) + CA (built-in) \\
    BAVE     & 62 & 62 & 3 clips, $\sim$15\,min & CA (ours) \\
    \bottomrule
  \end{tabular}%
  }
  \vspace{2pt}
  \small Note: For SEED and BAVE, annotators were recruited independently of EEG participants.
\end{table}

\subsection{Feature Extraction}

For each 1\,s segment $i$ with $s$ sample points $x(1),\dots,x(s)$ (where $s$ equals the sampling rate, e.g., 200\,Hz $\to$ $s{=}200$): First-order Difference (FD) computes $V_{FD}(i) = \frac{1}{s}\sum_{j=1}^{s-1} |x(j+1)-x(j)|$~\cite{yu_review_2019}. Raw computes $V_O(i) = \frac{1}{s}\sum_{j=1}^s |x(j)|$. Differential Entropy (DE)~\cite{duan_differential_2013} is extracted across five frequency bands per segment ($\delta$ 1--4\,Hz, $\theta$ 4--7\,Hz, $\alpha$ 8--13\,Hz, $\beta$ 14--29\,Hz, $\gamma$ 30--47\,Hz): $V_{DE}(i) = \frac{1}{2}\log(2\pi e \sigma_i^2)$, where $\sigma_i^2$ is the signal variance of segment $i$ after bandpass filtering. FD serves as the primary input feature.

\begin{figure}[h]
  \centering
  \includegraphics[width=\columnwidth]{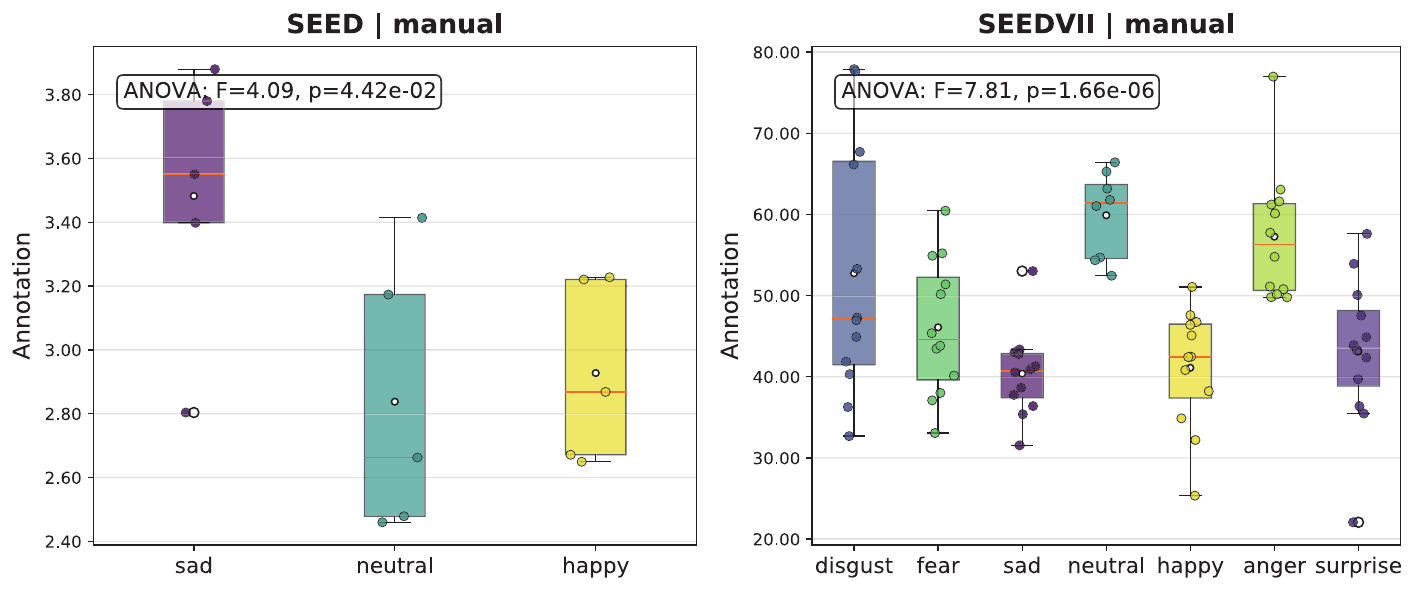}
  \caption[Valence-dependent arousal differences in manual annotations]{Valence-dependent differences in manually annotated continuous arousal (SEED and SEED-VII). Continuous arousal annotations (1\,Hz, per-clip averaged across $\geq$20 raters) are grouped by valence category (positive, neutral, negative). One-way ANOVA reveals significant valence-dependent arousal modulation: SEED ($F{=}4.09$, $p{=}4.42{\times}10^{-2}$, sad elicits highest arousal) and SEED-VII ($F{=}7.81$, $p{=}1.66{\times}10^{-6}$, disgust and anger peak). The significant valence-dependent structure confirms that manual annotations carry meaningful emotion information and serve as a valid behavioral reference for downstream neural analyses.}
  \label{fig:valence_anno}
\end{figure}

\begin{figure*}[htbp!]
  \centering
  \includegraphics[width=\textwidth]{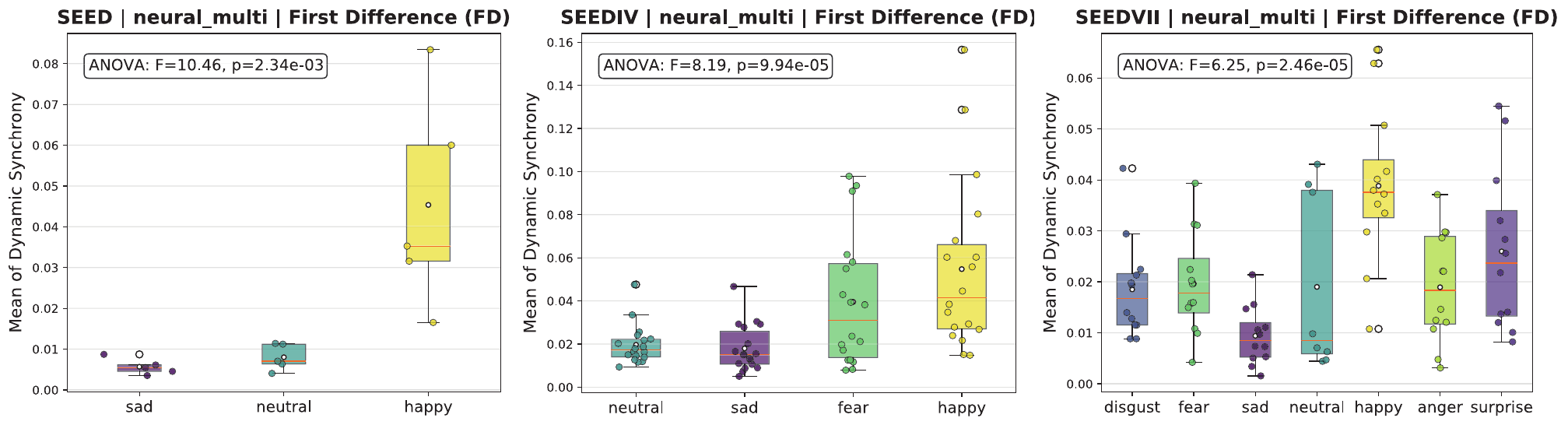}
  \caption[Valence-dependent DNS differences across datasets]{Valence-dependent DNS differences across datasets. DNS is computed using FD features with CorrCA Component~0 at window size 30\,s. Each panel displays the mean DNS per valence category with individual clip-level data points (dots) and box-and-whisker overlay. One-way ANOVA confirms significant valence-dependent DNS modulation: SEED ($F{=}10.46$, $p{=}2.34{\times}10^{-3}$), SEED-IV ($F{=}8.19$, $p{=}9.94{\times}10^{-5}$), SEED-VII ($F{=}6.25$, $p{=}2.46{\times}10^{-5}$), all $p{<}0.003$. Positive emotions elicit consistently higher synchrony than negative or neutral conditions across datasets, indicating that DNS inherits emotion-relevant structure from shared neural processing and is not merely a domain-general engagement signal.}
  \label{fig:valence_diff}
\end{figure*}

\subsection{CorrCA and DNS Computation}

CorrCA~\cite{dmochowski_correlated_2012} finds spatial filters $\mathbf{w}$ that maximize ISC across subjects. For $N$ subjects, we enumerate all $P{=}N(N{-}1)/2$ unordered subject pairs $\{(p_{i1},p_{i2})\}_{i=1}^P$ and form aggregated within- and cross-subject covariance matrices: $\mathbf{R}_{11}{=}\frac{1}{PT}\sum_i \mathbf{X}_{p_{i1}}\mathbf{X}_{p_{i1}}^T$, $\mathbf{R}_{22}{=}\frac{1}{PT}\sum_i \mathbf{X}_{p_{i2}}\mathbf{X}_{p_{i2}}^T$, $\mathbf{R}_{12}{=}\frac{1}{PT}\sum_i \mathbf{X}_{p_{i1}}\mathbf{X}_{p_{i2}}^T$, where $T$ is the number of time samples and $\mathbf{X}_n$ is subject $n$'s spatiotemporal data. To ensure $\mathbf{R}_{11}{=}\mathbf{R}_{22}$, each pair is duplicated with reversed order. The optimization $\max_{\mathbf{w}} \frac{\mathbf{w}^T \mathbf{R}_{12} \mathbf{w}}{\sqrt{\mathbf{w}^T \mathbf{R}_{11} \mathbf{w} \cdot \mathbf{w}^T \mathbf{R}_{22} \mathbf{w}}}$ reduces to a generalized eigenvalue problem $(\mathbf{R}_{11}{+}\mathbf{R}_{22})\mathbf{w}{=}\lambda\mathbf{R}_{12}\mathbf{w}$, regularized by retaining the top 10 dimensions of $\mathbf{R}_{11}{+}\mathbf{R}_{22}$. The leading component (Component~0) captures the dominant group-synchronous EEG pattern. DNS is computed in sliding windows ($W \in \{2,\ldots,40\}$\,s, 1-second step, $\geq$50\% overlap), with CorrCA weights learned from full-clip data and applied within each window. CorrCA operates at the group level; DNS values represent stimulus-driven group-level neural responses.

\subsection{Statistical Analysis}

 \textbf{Annotation Reliability.} Inter-rater consistency of manual annotations was assessed via split-half reliability: raters were randomly partitioned into two halves across 100 iterations, and the Pearson correlation between the two resulting mean annotation time series was computed per clip. \textbf{Valence.} Valence-dependent differences in annotations and DNS were tested using one-way ANOVA with $F$-statistic. \textbf{DNS--Arousal Coupling.} Per-clip Pearson $r$ between the DNS time series and the arousal (or arousal-derivative) time series was computed, then averaged across clips within each condition to produce a single summary statistic; 95\% confidence intervals were obtained via bootstrapping. \textbf{Circular Shift Null.} To test whether observed DNS-arousal correlations exceed chance-level temporal alignment, the DNS sequence was subjected to random-step circular shifting (1,000 iterations, FDR-corrected)~\cite{perez_conscious_2021}, disrupting any genuine temporal coupling while preserving the autocorrelation structure of both signals. \textbf{Block Permutation.} To rule out spurious associations driven by dataset-level confounds, DNS and arousal sequences were randomly re-paired across clips treated as independent blocks (10,000 permutations)~\cite{nastase_measuring_2019}, with significance assessed against the 95th percentile of the null distribution. \textbf{Subject-Split Validation.} To verify that DNS is a reliable group-level phenomenon, subjects were randomly split into two equal groups (100 splits); the full CorrCA pipeline was applied independently to each group, inter-group DNS consistency was assessed via Pearson $r$, and the DNS--arousal association was replicated within each subgroup.


\section{Experimental Results}

We present results around three core findings: DNS as a valid emotion-sensitive signal, DNS preferentially capturing arousal change rate, and the systematic parameterization and validation of the DNS-arousal coupling.

\subsection{DNS as a Valid Emotion-Sensitive Signal}
\label{sec:valence}

\textbf{Annotation Reliability and Valence Validation.}
Before examining neural synchrony, we first validate that the manual annotations are reliable and capture meaningful emotion structure. Split-half across 100 rater splits confirms high inter-rater consistency (SM Fig.~S1): SEED ($r{=}0.836$), SEED-VII ($r{=}0.960$), BAVE ($r{=}0.760$).

As an additional behavioral check, we verify valence-dependent arousal differences in manual annotations, a well-established phenomenon in affective science~\cite{sharma_continuous_2020,zhang_rcea_2020}. One-way ANOVA confirms significant valence-dependent arousal differences (Fig.~\ref{fig:valence_anno}): SEED ($F{=}4.09$, $p{=}4.42{\times}10^{-2}$) and SEED-VII ($F{=}7.81$, $p{=}1.66{\times}10^{-6}$). In SEED, ``sad'' elicits the highest arousal; in SEED-VII, ``disgust'' and ``anger'' peak. These valence-dependent differences confirm that the annotations carry meaningful emotion information.

\textbf{DNS Captures Valence Information.}
Having validated the annotations, we ask whether DNS, computed purely from neural data, similarly discriminates valence categories. One-way ANOVA reveals significant DNS differences across valence categories (Fig.~\ref{fig:valence_diff}): SEED ($F{=}10.46$, $p{=}2.34{\times}10^{-3}$), SEED-IV ($F{=}8.19$, $p{=}9.94{\times}10^{-5}$), SEED-VII ($F{=}6.25$, $p{=}2.46{\times}10^{-5}$), all $p{<}0.003$. Positive emotions elicit higher synchrony than negative or neutral conditions, suggesting that DNS inherits emotion-relevant structure from shared neural processing.

\subsection{DNS Captures Arousal Change Rate, Not Static Intensity}
\label{sec:derivative}

DNS, by construction, captures moment-to-moment fluctuations in inter-subject coupling. Prior fMRI work has shown that high-arousal states enhance ISC~\cite{nummenmaa_emotions_2012}, motivating arousal as a natural target for DNS. However, the precise form of coupling is not obvious a priori: DNS could reflect momentary arousal intensity (raw values), the rate at which arousal shifts (first-order derivative), or some combination of both. We therefore explored which operationalization best aligns with DNS across datasets and modalities. The results converge on a consistent answer: DNS preferentially tracks arousal change rate, not static intensity.

\begin{figure*}[t]
 \centering
 \includegraphics[width=\textwidth]{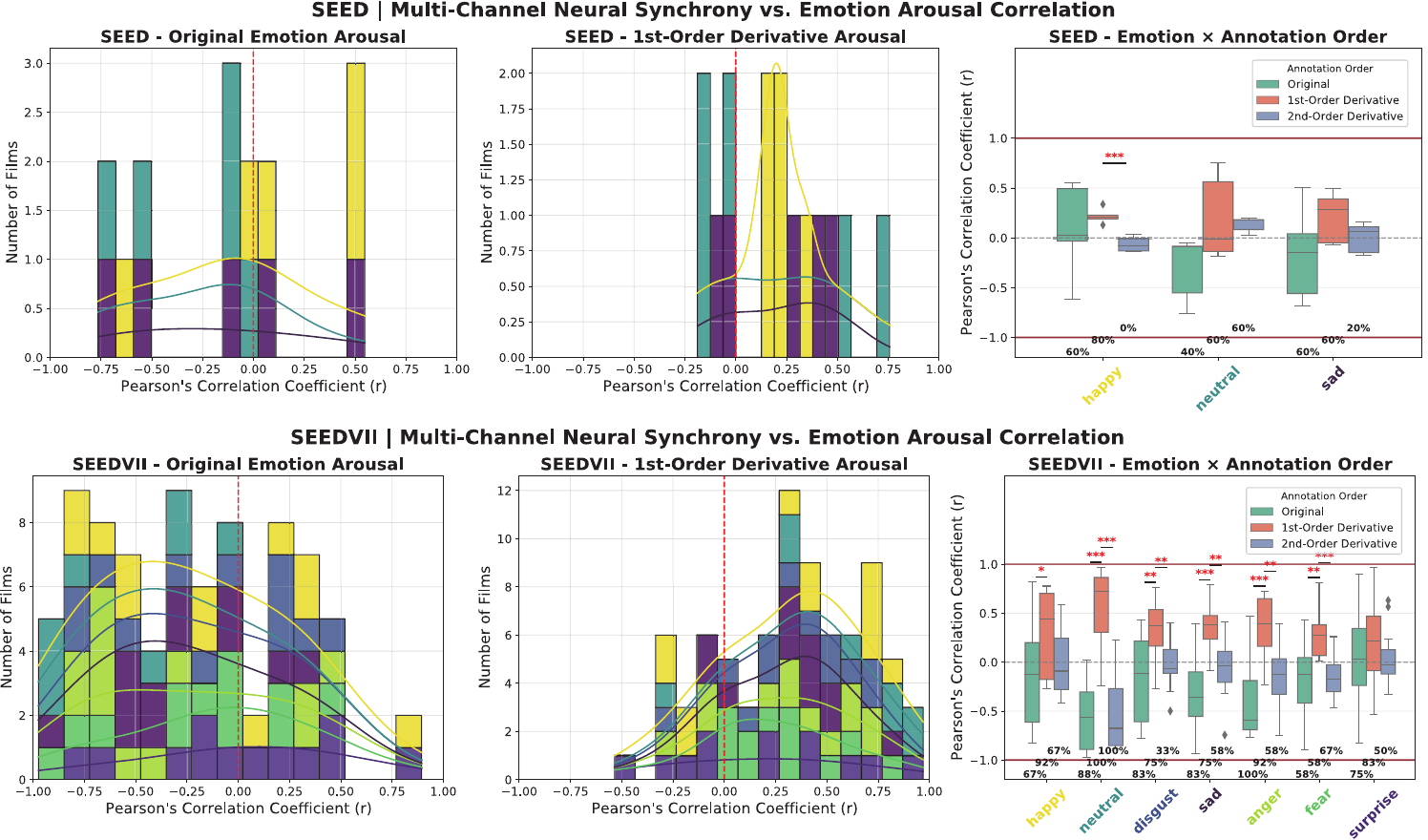}
 \caption[Cross-dataset DNS-arousal correlation distributions]{Cross-dataset DNS-arousal correlation distributions. DNS is computed using FD features with CorrCA Component~0 (window size 30\,s). Each row corresponds to one dataset (SEED top, SEED-VII bottom). \textbf{First two columns:} Per-clip Pearson correlations ($r$) between DNS and three arousal operationalizations, raw intensity ($r_{\text{raw}}$), first-order derivative ($r_{\text{deriv}}$), and second-order derivative ($r_{\text{2nd}}$), shown as stacked histograms color-coded by valence category (positive, neutral, negative), with overlaid density curves and a red dashed line at $r{=}0$. \textbf{Third column:} Box-and-whisker plots of per-clip $r$ values grouped by valence category and derivative order. Asterisks mark FDR-corrected paired comparisons between $r_{\text{deriv}}$ and the other two orders: *$p{<}0.05$, **$p{<}0.01$, ***$p{<}0.001$. Numbers beneath each box indicate the proportion of films within that valence category whose DNS--arousal correlation reaches FDR-corrected significance. In SEED-VII, $r_{\text{deriv}}$ distributions concentrate more strongly in the positive range and achieve significantly higher proportions of significant films than $r_{\text{raw}}$ and $r_{\text{2nd}}$ (FDR $p{<}0.001$). The consistent $r_{\text{deriv}}$ advantage supports the conclusion that neural synchrony is mechanistically more tightly coupled to arousal change rate than to static arousal intensity.}
  \label{fig:cross_dataset}
\end{figure*}

\begin{figure*}[t]
 \centering
 \includegraphics[width=\textwidth]{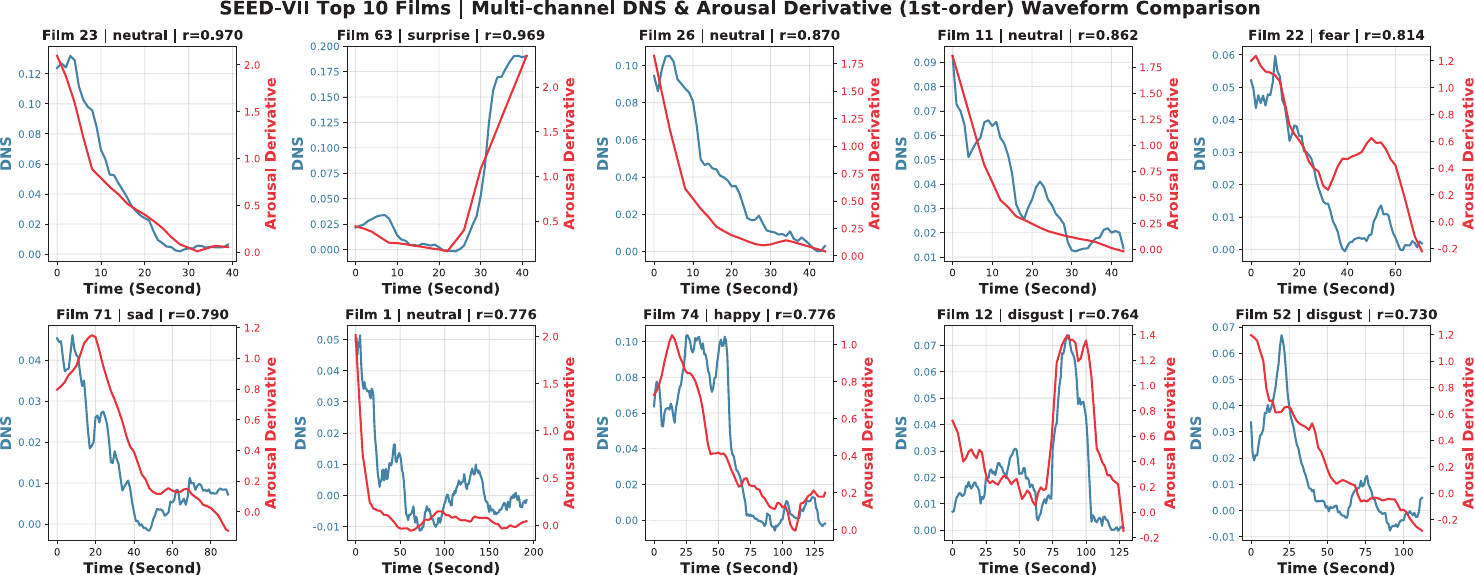}
 \caption[DNS and arousal derivative waveforms for top-10 SEED-VII films]{DNS (blue) vs.\ arousal derivative (red) for top-10 SEED-VII films. DNS is computed using FD features with CorrCA Component~0 (window size 30\,s, 1-second step). Each panel shows the normalized DNS time course (blue) overlaid with the per-clip  arousal derivative (red, first-order difference of manual annotations, resampled to 1\,Hz) for one film. The x-axis represents time in seconds; y-axes are individually scaled to highlight within-film fluctuations. It illustrates the temporal coupling between group-level neural synchrony and emotional change rate at the single-film level.}
  \label{fig:seedvii_top10}
\end{figure*}

\textbf{Cross-Dataset Consistency.} In SEED and SEED-VII (Fig.~\ref{fig:cross_dataset}), $r_{\text{deriv}}$ distributions concentrate more strongly in the positive range. SEED-VII $r_{\text{deriv}}$ achieves significantly higher $r$ than $r_{\text{raw}}$ and $r_{\text{2nd}}$ (FDR $p{<}0.001$). Waveform comparisons for top-10 SEED-VII films confirm synchronous fluctuations (Fig.~\ref{fig:seedvii_top10}).

\textbf{BAVE Multi-Modal Evidence.} $r_{\text{deriv}}$ significantly exceeds $r_{\text{raw}}$ for audio ($r_{\text{deriv}}{=}0.294$, $p{=}9.23{\times}10^{-21}$ vs.\ $r_{\text{raw}}{=}0.189$, $p{=}3.12{\times}10^{-9}$) and audiovisual ($r_{\text{deriv}}{=}0.253$, $p{=}7.79{\times}10^{-15}$ vs.\ $r_{\text{raw}}{=}0.220$, $p{=}1.41{\times}10^{-11}$) modalities (SM Fig.~S2,~S3). Video-only shows comparable correlations ($r_{\text{raw}}{=}0.203$, $p{=}4.22{\times}10^{-11}$, $r_{\text{deriv}}{=}0.114$, $p{=}2.27{\times}10^{-4}$), with lower arousal likely reflecting the reduced engagement of silent films; nonetheless, multiple arousal-derivative peaks align visibly with DNS fluctuations (SM Fig.~S3).

The absolute $r$ values are modest ($r_{\text{deriv}}$ mean 0.05--0.20), as expected for group-level EEG signals aggregating across heterogeneous individual responses. Despite their magnitude, these correlations are consistently significant across datasets and modalities. The $r_{\text{deriv}} > r_{\text{raw}}$ advantage holds across SEED, SEED-VII and all three BAVE clips, supporting the conclusion that neural synchrony is mechanistically more tightly coupled to arousal change rate than to static arousal intensity.

\subsection{Parameterizing and Validating the DNS-Arousal Coupling}
\label{sec:parameterize}

The preceding sections establish that a systematic DNS-arousal coupling exists. But is it stable, or does it depend on arbitrary methodological choices? Three factors demand characterization: window size (which determines the temporal grain of synchrony estimation), temporal alignment (whether DNS leads or lags arousal), and feature selection (which EEG representation best captures emotion-related synchrony). All analyses in this section use FD with Component~0, the empirically validated default established above.

\subsubsection{Window Size Modulates DNS Sensitivity}

Window size has been treated as a fixed hyperparameter in prior EEG-ISC work~\cite{dmochowski_correlated_2012,ding_interbrain_2021}, with no investigation of its impact on emotion-related coupling. We provide the first comprehensive evaluation of how window size modulates both valence discrimination and arousal coupling.

For valence detection (SM Fig.~S4), medium windows (10--30\,s) yield strongest F-values across SEED-IV and SEED-VII; SEED shows a U-shaped pattern. For arousal correlations (SM Fig.~S5), $r_{\text{deriv}}$ increases with window size (SEED, SEED-VII) or peaks at 10--25\,s (BAVE), while $r_{\text{raw}}$ remains flat. The $r_{\text{deriv}} - r_{\text{raw}}$ gap widens with window size.

Taken together, these sweeps reveal that window size is not a trivial hyperparameter: its choice meaningfully modulates both whether DNS detects valence information and how strongly it tracks arousal change. The convergence across datasets on 10--30\,s as a consistently effective range points to a likely mechanistic explanation: arousal fluctuations in naturalistic stimuli unfold on timescales of seconds to tens of seconds, driven by narrative pace and scene transitions. Windows shorter than 10\,s capture transient fluctuations with insufficient inter-subject alignment; windows beyond 30\,s begin to average across heterogeneous emotional segments, attenuating the temporal specificity of the coupling. The 10--30\,s range thus represents a regime where temporal granularity and signal-to-noise ratio are jointly optimized.

\subsubsection{Temporal Alignment}

\begin{figure*}[t]
 \centering
 \includegraphics[width=\textwidth]{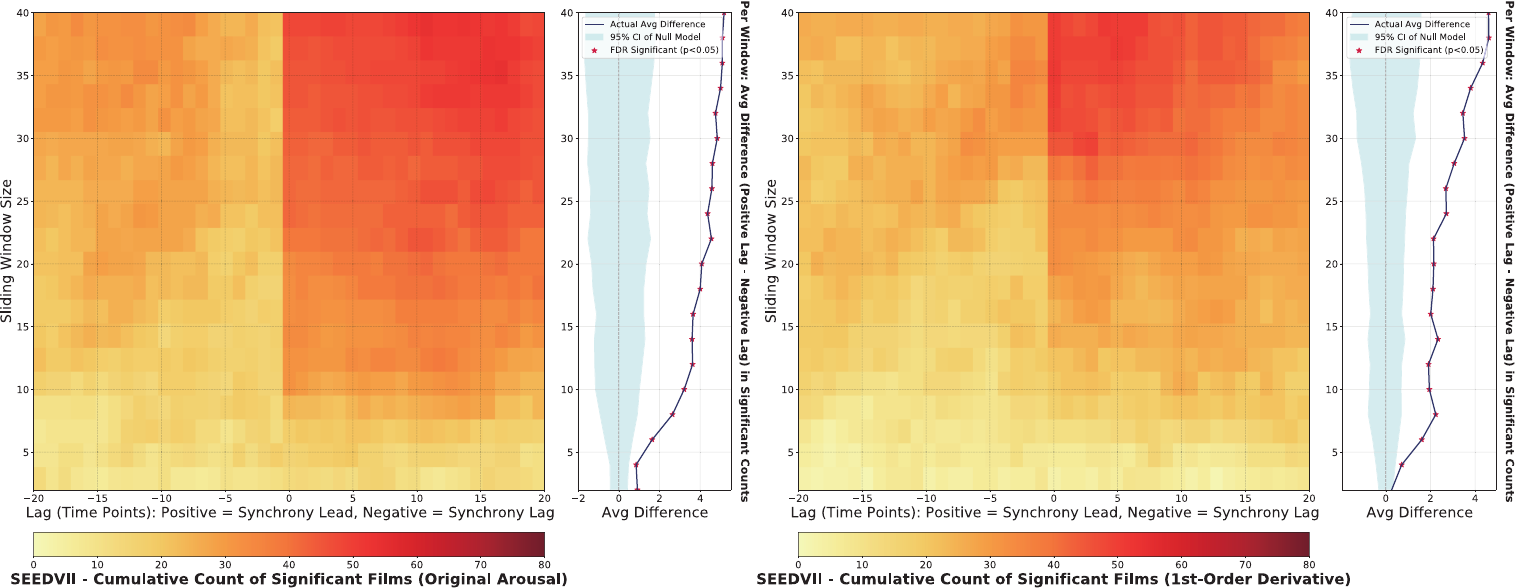}
 \caption[Window-lag optimization for DNS-arousal coupling (SEED-VII)]{Window-lag optimization for DNS--arousal coupling (SEED-VII). DNS is computed using FD features with CorrCA Component~0 across a grid of window sizes (2--40\,s) and temporal lags ($-$20 to $+$20 steps, 1-second step). Each cell in the heatmap represents the number of films in SEED-VII for which DNS--arousal (first-order derivative or original) correlation is statistically significant after FDR correction at the corresponding window size (x-axis) and time lag (y-axis). Red asterisks mark parameter pairs where the observed significant-film count exceeds the 95th percentile of a permutation-based null distribution ($p{<}0.05$). Significant associations concentrate at positive lags (0--10 steps) with medium windows (15--30\,s), indicating that DNS tends to lead annotated arousal, consistent with multi-stage emotion processing in which neural representations emerge before subjective awareness.}
 \label{fig:param_opt}
\end{figure*}

Beyond correlation strength, the temporal relationship carries practical and theoretical implications: does neural synchrony lead or lag the annotated arousal signal? A consistent lead would motivate predictive uses; a lag would suggest post-hoc reflection. Significant associations concentrate at positive lags (0--10 steps) with medium windows (15--30\,s) across SEED-VII (Fig.~\ref{fig:param_opt}), SEED (SM Fig.~S6), and BAVE (SM Fig.~S7). We note two caveats in interpreting this positive lag. Theoretically, it is compatible with multi-stage emotion processing, in which neural representations emerge before subjective awareness~\cite{meuleman_nonlinear_2013}. Methodologically, post-hoc annotation inherently introduces response latency, and window centering can shift the effective temporal reference point. Disentangling these contributions from genuine neural anticipation requires future work with precisely time-stamped stimulus events.

\textbf{Feature and Component Selection.}
In this section, we systematically evaluate the impact of feature type and component order on DNS quality. FD and DE are established features in EEG-based emotion recognition~\cite{zheng_investigating_2015,duan_differential_2013}, motivating their adoption for inter-subject synchrony analysis. FD with Component~0, used throughout all preceding analyses, yields the highest global synchrony across all datasets (SM Fig.~S8), and we further evaluate DE (separated by frequency band) and raw amplitude to systematically benchmark alternative configurations.

Among alternatives, DE-gamma with Component~0 achieves comparable $r_{\text{deriv}}$ values (SM Fig.~S9). FD and DE-gamma time courses are qualitatively consistent (SM Fig.~S10). Lower-frequency DE bands yield systematically weaker coupling, consistent with gamma-band specificity of emotion-related synchrony~\cite{zheng_investigating_2015}. Higher-order components (1--4) capture progressively weaker shared variance across all feature types. DE-gamma remains a viable alternative when gamma-band information is specifically relevant.

\textbf{Robustness Validation.}
To assess whether the observed DNS-arousal coupling could arise from temporal autocorrelation or chance alignment rather than genuine coupling, we employ two complementary resampling strategies: within-dataset block permutation and subject-split validation.

\textbf{Block Permutation.} Treating each clip as an independent block, we randomly re-pair DNS and arousal sequences across films (10,000 permutations). At window 30\,s, $r_{\text{deriv}}$ significantly exceeds the 95th null percentile for both SEED-VII ($r_{\text{obs}}{=}0.329$, $r_{95}{=}0.209$, $p{<}0.001$) and SEED ($r_{\text{obs}}{=}0.208$, $r_{95}{=}0.178$, $p{=}0.019$). BAVE is excluded from block permutation (only 3 films, insufficient for reliable null estimation). $r_{\text{raw}}$ is not significant for any dataset.

\begin{table}[tb]
  \centering
  \caption[Block permutation test for DNS-arousal coupling]{Block permutation test for DNS--arousal coupling at window 30\,s. DNS and arousal sequences are randomly re-paired across films (10,000 permutations). $r_{\text{obs}}$: observed mean. $r_{95}$: 95th null percentile. Bold: $r_{\text{obs}} > r_{95}$ ($p{<}0.05$). BAVE excluded (3 films).}
  \label{tab:block_perm}
  \resizebox{\columnwidth}{!}{%
  \begin{tabular}{lcccc}
    \toprule
    \textbf{Dataset} & \textbf{$r_{\text{raw}}$} & \textbf{$r_{\text{95}}$} & \textbf{$r_{\text{deriv}}$} & \textbf{$r_{\text{95}}$} \\
    \midrule
    SEED     & -0.131 & -0.077 & \textbf{0.208} & 0.178 \\
    SEED-VII & -0.231 & -0.158 & \textbf{0.329} & 0.209 \\
    \bottomrule
  \end{tabular}%
  }
\end{table}

\textbf{Subject-Split Validation.} Across 100 random splits at window 30\,s, inter-group DNS consistency reaches $r{=}0.383$ (SEED), $r{=}0.413$ (SEED-VII), $r{=}0.698$ (BAVE). Both-significant ratios$\,$, the percentage of splits where both subject groups independently yield significant DNS-arousal coupling$\,$, reach 50.2\%, 58.2\%, and 59.3\% for SEED, SEED-VII, and BAVE, respectively. 


\section{Discussion and Conclusion}

This study establishes group-level EEG DNS as an empirically validated signal for continuous emotional arousal dynamics. Through systematic evaluation across four datasets, comprehensive parameter sweeps, and conservative validation, we demonstrate that DNS captures emotion-specific information, preferentially reflects arousal change rate, and that this coupling is systematically governed by parameters and robust under controls. In particular, larger window sizes (10--30\,s), positive lags, and the default configuration of FD with the dominant CorrCA component consistently enhance the DNS-arousal coupling, and both block permutation and subject-split validation confirm the robustness of this coupling.

\subsection{From ISC to Continuous Emotion Quantification}

The central contribution is demonstrating that group-level EEG neural synchrony, extensively validated in cognitive neuroscience for shared narrative processing and engagement~\cite{hasson_intersubject_2004,dmochowski_audience_2014,nastase_measuring_2019}, can be systematically exploited for continuous emotion quantification. This extends ISC from cognitive into the affective domain and into the challenging regime of continuous dynamics.

The finding that DNS tracks arousal change rate has theoretical and practical significance. Theoretically, it aligns with evidence that the brain is more sensitive to changing stimuli~\cite{somervail_waves_2020}, suggesting that DNS operates on the same principle: its moment-to-moment fluctuations inherently reflect the brain's sensitivity to dynamic shifts in emotional state. In this light, DNS does not merely echo static arousal labels but reveals a distinct temporal dimension, the rate at which collective neural representations reorganize as emotions rise and fall. Practically, this offers a complementary lens for emotion quantification: rather than depending on static intensity judgments that annotation paradigms are designed to deliver, DNS captures the temporal dynamics of emotional experience directly from group-level neural activity, providing access to a signal stream whose temporal granularity matches the brain's own processing timescale.

The positive time lag (DNS leading annotations) suggests a temporal structure in which group-level neural synchrony shifts before annotated arousal ratings change. This finding is compatible with multi-stage views of emotion processing~\cite{meuleman_nonlinear_2013,goldstein_shared_2022}, where early neural representations precede later subjective experience. However, we interpret this lag conservatively as systematic temporal alignment rather than mechanistic causality, pending future studies that explicitly control for annotation response latency and stimulus-level timing factors.

\subsection{Toward Annotation-Efficient Emotion Quantification}

Continuous emotion annotation is a persistent bottleneck in affective computing. Current practice requires dozens of trained raters to view every stimulus and provide moment-by-moment ratings, a process that is time-consuming, labor-intensive, and fundamentally non-scalable. Each hour of stimulus material demands hundreds of person-hours of annotation effort. When stimuli, experimental paradigms, or recording conditions change, the entire annotation pipeline must be repeated, creating a recurring cost that constrains both research throughput and real-world deployment.

DNS offers a fundamentally different operating model. In the conventional approach, annotation is an active cognitive task: raters must continuously introspect on their fluctuating emotional state and translate it into moment-by-moment numerical ratings. This demands sustained attentional effort, introduces dual-task interference with the stimulus experience itself, and limits the scale of data collection to what a small number of trained annotators can sustain. DNS replaces this active labeling with passive viewing. Participants simply watch the stimulus material while EEG is recorded; the neural synchrony that naturally emerges across a group serves as the signal. No introspection, no rating interface, no cognitive load beyond attending to the stimulus. Once the DNS-arousal coupling is validated on a modest set of annotations, a one-time calibration, subsequent groups contribute usable arousal dynamics simply by watching. In effect, the annotation cost is paid once and amortized across all studies and participants thereafter.

This shifts the bottleneck from annotation throughput, hundreds of person-hours per dataset, to EEG acquisition, which is increasingly accessible with portable dry-electrode systems and requires only passive viewing from participants. The practical benefit is clearest in longitudinal monitoring, adaptive interface design, and large-scale content evaluation, where repeated per-subject labeling is simply infeasible. The methodological guidelines established above, moderate windows (10--30\,s), FD with Component~0, and awareness of the positive lag, provide practitioners with an operational starting point that does not require per-study parameter re-tuning. Convergence across SEED-VII and BAVE strengthens generalizability beyond a single dataset or annotation protocol.

\subsection{Limitations}

\textbf{Group-level nature.} DNS is inherently group-level; subject-split validation confirms cross-subset reproducibility, but DNS should not be interpreted as an individual-level decoder. Bridging to individual prediction via covariates is future work.

\textbf{Low-level confounds.} Uncontrolled audiovisual features (e.g., luminance and motion energy) may contribute to DNS. Mitigating factors: consistent valence effects across qualitatively different stimulus materials, and BAVE's multi-modal design showing the derivative advantage for audio and audiovisual stimuli with different acoustic and visual properties. Parametric stimulus manipulation is needed.

\textbf{Ecological validity.} Our evaluation focuses on passive video viewing, a common real-world activity; whether DNS-arousal coupling extends to additional contexts such as social interaction or free-living settings warrants further study.

In conclusion, group-level EEG neural synchrony, when systematically parameterized and validated, provides a principled signal for continuous emotional arousal quantification. By establishing empirical foundations and methodological guidelines, we take a step toward annotation-efficient continuous arousal quantification.

\bibliographystyle{plainnat}
\bibliography{aaai2027}

\end{document}